\documentclass[twoside,11pt]{article}
\usepackage{jair, theapa, rawfonts}
\usepackage{adjustbox}
\usepackage{algorithm}
\usepackage{algorithmic}
\usepackage{setspace}
\usepackage{bm}
\usepackage{amssymb}
\usepackage{latexsym}
\usepackage{graphicx}
\usepackage{framed,multirow}
\usepackage{xcolor}
\usepackage[lofdepth,lotdepth]{subfig}

\jairheading{1}{1993}{1-15}{6/91}{9/91}
\ShortHeadings{Attributed Relational SIFT-based Regions Graph (ARSRG): concepts and applications}
{Manzo}
\firstpageno{1}

\begin{document}

\title{Attributed Relational SIFT-based Regions Graph (ARSRG): concepts and applications}

\author{\name Mario Manzo \email mmanzo@unior.it \\
       \addr Information Technology Services, University of Naples ``L'Orientale'',\\
       80121 Naples, Italy
       }


\maketitle

\begin{abstract}
Graphs are widely adopted tools for encoding information. Generally, they are applied to disparate research fields where data needs to be represented in terms of local and spatial connections. In this context, a structure for ditigal image representation, called \textbf{Attributed Relational SIFT-based Regions Graph} (\textbf{ARSRG}), previously introduced \cite{manzo2013attributed,manzo2014novel,manzo2019bag,manzo2019kgearsrg}, is presented. \textbf{ARSRG} has not been explored in detail in previous works and for this reason the goal is to investigate unknown aspects. The study is divided into two parts. A first, theoretical, introducing formal definitions, not yet specified previously, with purpose to clarify its structural configuration. A second, experimental, which provides fundamental elements about its adaptability and flexibility regarding different applications. The theoretical vision combined with the experimental one shows how the structure is adaptable to image representation including contents of different nature. 


\end{abstract}

\section{Introduction}
\label{Introduction}

Among issues related human vision the processing of visual complex entities is one of most important. The processing of information is often based on local-to-global or global-to-local connections
\cite{love1999structural}. Local-to-global concept concerns the transitions from local details of scene to global configuration,
while global-to-local works in the reverse order, from global configuration towards the details. For example an algorithm
for face recognition, which use local-to-global approach, starts eyes, nose and ears recognition, and finally brings to face
configuration. Differently, a global-to-local algorithm first identifies the face that leads to the identification of details (eyes, nose and ears). During the task of human recognition global configuration of a
scene plays a key role, especially when subjects see the images for a short duration of time. Also, humans leverage local information in effect way to recognize scene categories. Theories
of higher-level visual perception split individual elements
at the local level and global objects, for which the information on
many local components are perceptually grouped \cite{Koffka}. Graphs are frequently  adopted to represent information in terms of
nodes and edges, where relations among data must be highlighted and generally occur in raw
form. Computer Vision, Pattern Recognition and many other fields
benefit from data graph representations and related manipulation
algorithms. Specifically, in the Image Processing field, graphs are
used to represent digital images in many ways. Standard approach concerns partitioning of the image into dominant disjoint regions,  where local and spatial features are respectively nodes
and edges. Local features describe intrinsic properties of regions
(such as shape, colors, texture), while spatial features provide
topological information about neighborhood.  Image representation is one of the crucial steps for systems working
in the Image Retrieval field. Modern Content Based Image Retrieval (CBIR) systems consider
essentially the image basic elements (colors, textures, shapes and
topological relationships) extracted from the entire image, in order
to provide an effective representation. Through the analysis of
these elements, compositional structures are produced. Other
systems, called Region Based Image Retrieval \cite{liu2007survey}
(RBIR), focus their attention on specific image regions instead of the entire content to extract features. In this paper, a graph
structure for image representation, called \textbf{Attributed
Relational SIFT-based Regions Graph} (\textbf{ARSRG}), is described, analyzed and discussed with reference to previous works \cite{manzo2013attributed,manzo2014novel,manzo2019bag,manzo2019kgearsrg}. 
In particular, new definitions and properties arising from the detailed analysis of the structure are introduced. Finally, through a wide experimental phase, how the structure is adaptable to different types of application contexts is shown. The paper is organized as follows: section \ref{relatedwork} includes related research about graph based image representation including Scale Invariant Feature Transform (SIFT) \cite{Lowe}. Sections \ref{ARSRGdescr}, \ref{defs}, \ref{props} are dedicated to \textbf{ARSRG} description, definitions and properties. Experimental results and conclusions are respectively reported in section \ref{res} and section \ref{conc}.

\section{Related work}
\label{relatedwork}

The literature reports many approaches which combine local and spatial information arising from SIFT features. Commonly, a graph structure encodes information about keypoints located in a certain position of image. Nodes represent SIFT descriptors, while edges describe spatial relationships between different keypoints.

In \cite{SanromaAlquezarSerratosa} a graph $G_{1}$ represents a set of SIFT keypoints from the image $I_{1}$ and is defined as

\begin{equation} G_{1}=(V_{1},M_{1},Y_{1})\end{equation}

where $v_{\alpha} \in V_{1}$ is a node associated to a SIFT keypoint
with position $(p_{1}^{(\alpha)},p_{2}^{(\alpha)})$, $y_{\alpha} \in
Y_{1}$ is the SIFT descriptor attached to node $v_{\alpha}$ and
$M_{1}$ is the adjacency matrix. If $M_{1\;\alpha\beta}=1$ the nodes
$v_{\alpha}$ and $v_{\beta}$ are adjacent, $M_{1\;\alpha\beta}=0$
otherwise.

In \cite{SanromaManchoSerratosa2} authors combine local information of SIFT features with global geometrical
information in order to estimate a robust set of features-matches. These information are encoded using a graph structure

\begin{equation}G_{0}=(V_{0},B,Y)\end{equation}

where $v \in V_{0}$ is a node associated to a SIFT keypoint, $B$ is the adjacency matrix, $B_{v,v'}=1$ if the nodes $v$ and $v'$ are connected $B_{v,v'}=0$ otherwise, while $y_{v} \in Y$ is the SIFT descriptor associated to node $v$. 

In \cite{DuchenneJoulinPonce} nodes are associated to $N$ image
regions related to an image grid, while edges connect each node with its four neighbors. Basic elements are not pixels but regions
extended in the $x$ (horizontal) and $y$ (vertical) directions. The
nodes are identified using their coordinates on the grid. The
spatial information associated to nodes are indices
$d_{n}=(x_{n},y_{n})$. Also, a feature vector $F_{n}$ is associated
with the corresponding image region and, then, to node. The image is
divided into overlapping regions of $32$ $\times$ $32$ pixels. Four
$128$-dimensional SIFT descriptors, for each region, are extracted
and concatenated. 


In \cite{ChoLee} the graph based image representation includes SIFT features, MSER \cite{MatasChumUrbanPajdla}
and Harris-Affine \cite{MikolajczykSchmid2}. Given two graphs $G^{P}=(V^{P},E^{P},A^{P})$ and $G^{P}=(V^{Q},E^{Q},A^{Q})$, representing images $I^{P}$ and $I^{Q}$, $V$ is the set of nodes, image features extracted, $E$ the set of edges, features spatial relations, and $A$ the set of attributes, information
associated to features extracted. 

In \cite{LeeChoLee} SIFT features are combined in form of hyper-graph. A hyper-graph $G=(V,E,A)$ is composed of nodes $v \in V$, hyper-edges $e \in E$, and attributes $a \in A$ associated with the hyper-edges. A hyper-edge $e$ encloses a subset of nodes with size $\delta(e)$ from $V$, where $\delta(e)$ represents the order of an hyper-edge. 

In \cite{RevaudLavoueArikiBaskurt} an approach to $3$D objects recognition is presented. Graph matching
framework is used in order to enable the utilization of SIFT
features and to improve robustness. Differently to standard methods,
test images are not converted into finite graphs through operations
of discretization or quantization. Then, continuous graph space is
explored in the test image at detection time. To this end, local
kernels are applied to indexing image features and to enable a fast
detection. 

In \cite{RomeroCazorla2} an approach to matching features problem with application of scene recognition and topological SLAM is proposed. For this purpose, the scene images are encoded using a particular data structure. Image representation is built through two steps: image segmentation using JSEG \cite{DengManjunath} algorithm and invariant feature extraction MSER and SIFT descriptors in a combined way. 

In \cite{XiaHancock} SIFT features based on visual saliency and
selected to construct object models are extracted. A Class Specific Hypergraph ($CSHG$) to model objects in compact way is introduced. The hypergraphs are built on different Delaunay graphs. Each one is created from a set of selected SIFT features using a single prototype image of an object. Using this approach, the object models can be represented through a minimum of object views.

In \cite{HoriTakiguchiAriki} a method for generic object recognition through graph structural expression using SIFT features is described. A graph structure is created using lines to connect SIFT keypoints. The graph is represented as
$G=(V,E,X)$ where $E$ represents the set of edges, $V$ is the set of vertices and $X$ the set of their associated labels, SIFT
descriptors. The node represents a keypoint detected by SIFT
algorithm and the associated label is the $128$-dimension SIFT
descriptor. The edge $e_{\alpha\beta} \in E$ connects two nodes
$u_{\alpha} \in V$ and $u_{\beta} \in V $. The graph is complete
when all keypoints extracted from the image are connected by edges.
Formally, the set of edges is defined as follows:

\begin{equation} E=\left\{e_{ij} \mid \forall i,j \frac{\parallel
p_{i}-p_{j}\parallel}{\sqrt{\sigma_{i}\sigma_{j}}} <
\lambda\right\}\end{equation}

where $p=(p_{x},p_{y})$ represents keypoint spatial coordinates,
$\sigma$ its scale, and $\lambda$ is a threshold value. An edge does
not exist when the value is greater than the threshold $\lambda$. In
this way, an extra edge is not created. This formulation of
proximity graph reduces the computation complexity and, at same
time, improves the detection performance.

In \cite{LuoQi} a median \emph{K-nearest-neighbor} (K-NN) graph
$G_{P}=(V_{P},E_{P})$ is built. A vertex $v_{i}$ for each of the $N$
points $p_{i}$ is created, with $V_{P}=v_{1},...,v_{N}$. Also, a
non-directed edge $(i,j)$ is created when $p_{j}$ is one of the $K$
closest neighbors of $p_{i}$ and $\parallel p_{i}-p_{j}\parallel
\leq \eta$. $\eta$ is the median of all distances between pairs of
vertices and is defined as:

\textbf{\begin{equation}\eta=median_{(l,m)\in V_{P} \times V_{P}}|| p_{l}-p_{m}||\end{equation}}

If there are not $K$ vertices that support the structure of $p_{i}$
then this vertex is completely disconnected until the end of the
K-NN graph construction. The graph $G_{P}$ has the $N \times N$
adjacency matrix $A_{P}$, where $A_{P}(i,j)=1$ when $(i,j) \in
E_{P}$ and $A_{P}(i,j)=0$ otherwise.

\section{Attributed Relational SIFT-based Regions Graph (ARSRG)}
\label{ARSRGdescr}

In this section \textbf{Attributed Relational SIFT-based Regions Graph (ARSRG)} is introduced based on two main steps: features extraction and graph construction. The first step consists of Regions of Interest (ROIs)
extraction from the image through a segmentation technique.
Connected components in the image are then identified with the aim
of building the \emph{Region Adjacency Graph (RAG)}
\cite{Tremeau:regions}, to encode spatial relations between image
regions. Simultaneously, SIFT \cite{Lowe} descriptors are extracted
from the original image, in order to ensure invariance to image
rotation, scaling, translation, illumination changes and projective
transforms. The second step consists in the construction of graph
structure. \textbf{ARSRG} is composed of three levels:
\emph{root node}, \emph{RAG nodes} and \emph{leaf nodes}. At first
level, the \emph{root node} represents the image and is linked to
all \emph{RAG nodes} at the second level. \emph{RAG nodes} encode
adjacency relationships between different image regions. Thus,
adjacent regions in the image are represented by connected nodes. In
addition, each \emph{RAG} node is connected with the \emph{Root
node} at the higher level. Finally, the \emph{leaf nodes} represent
the set of SIFT descriptors extracted from the image. At third
level, two types of configurations are provided: \emph{Region based}
and \emph{Region graph based}. In the \emph{Region based}
configuration, a keypoint is associated to a region based on its
spatial coordinates, whereas \emph{Region graph based} configuration
describes keypoints belonging to the same region connected by edges
(which encode spatial adjacency). Below, the steps of features
extraction and graph construction are described in detail.

\subsection{Features extraction}

\subsubsection{Region of interests (ROIs) extraction}

ROIs from the image through a segmentation algorithm
called JSEG \cite{DengManjunath} are extracted. JSEG performs segmentation through
two different steps: color quantization and spatial segmentation.
First step consists in a coarse quantization without degrading the
image quality significantly. In the second step, a spatial
segmentation directly on the class-map without taking
into account the color similarity of the corresponding pixel is performed.

\subsubsection{Labeling connected components}

The next step involves the labeling of connected components on the segmentation result. A connected component is an image region consisting of contiguous pixels of the same color. The process of connected components labeling of an image $B$ produces an output image $LB$ that contain labels (positive integers or characters). A label is a symbol naming an entity exclusively. Regions connected by the $4$-neighborhood and $8$-neighborhood will have the same label. Algorithm \ref{CCL} shows a version of connected components labeling.




\begin{algorithm}[!ht]
\caption{$Connected\;Components\;Labeling$} 
\label{CCL}
\begin{algorithmic}[1]
\REQUIRE $I$ - Image to Label; \ENSURE $I$ - Image Labeled;

\STATE m=0

\FOR{y=1:$I\_size\_y$}

\FOR{x=1:$I\_size\_x$}

\IF{I[i][j] == 0}

\STATE m=m+1 \STATE $Component\;Label(I,x,y,m)$

\ENDIF

\ENDFOR

\ENDFOR
\RETURN I
\end{algorithmic}
\end{algorithm}


\begin{algorithm}[!ht]
\caption{$Component\;Label$} \label{alg2}
\begin{algorithmic}[1]
\REQUIRE $I$ - Image to Label; $i,j$ - image index; $l$ - label;
\ENSURE $\oslash$;
\medskip

\IF{I[i][j] == 0}

\STATE I[i][j]=m

\STATE $Component\;Label(I,i-1,j-1,m)$

\STATE$Component\;Label(I,i-1,j,m)$

\STATE$Component\;Label(I,i-1,j+1,m)$

\STATE $Component\;Label(I,i,j-1,m)$

\STATE $Component\;Label(I,i,j+1,m)$

\STATE $Component\;Label(I,i+1,j-1,m)$

\STATE $Component\;Label(I,i+1,j,m)$

\STATE $Component\;Label(I,i+1,j+1,m)$

\ENDIF
\end{algorithmic}
\end{algorithm}

\subsubsection{Region Adjacency Graph (RAG) structure}
\label{RAGstr}

The \emph{Region Adjacency Graph (RAG)} \cite{Tremeau:regions} is adopted to build a graph based image representation located at second level of the \textbf{ARSRG}
structure. Based on image segmentation result, a region represents an elementary component of
the image. \emph{RAG} is built with reference to spatial relations
between regions. Two regions are defined to be adjacent if they
share the same boundary. In the \emph{RAG}, a node represents a
region, and an edge represents adjacency between two nodes. The
\emph{RAG} is defined as a graph $G=(V,E)$, where nodes are regions
in $V$ and edges $E$ identify the boundaries that connect them.
Moreover, the \emph{RAG} connectivity is invariant to translations
and rotations, which is a useful property for a high-level image
representation. In the algorithm \ref{RAGalg}, a pseudocode version
of \emph{RAG} algorithm is shown.

\begin{algorithm} [!ht]
\caption{$Region\;Adjacency\;Graph$} \label{RAGalg}
\begin{algorithmic}[1]
\REQUIRE $Labeled\_image$; \ENSURE $Graph$ $Structure$
$(Adjacency\_matrix)$;
\medskip

\STATE $Adjacency\_matrix=0$
\medskip

 \FOR{$pixel(i,j) \in Labeled\_image$}

\FOR{$pixel(x,y) \in 8-neighborhood$}

\IF{$pixel(i,j)\neq pixel(x,y)$}

\STATE $Adjacency\_matrix(pixel(i,j),pixel(x,y))=1$

\ENDIF

\ENDFOR

\ENDFOR

\RETURN $Adjacency\_matrix$
\end{algorithmic}
\end{algorithm}

\subsubsection{Scale Invariant Feature Transform (SIFT)}

SIFT \cite{Lowe} descriptors are extracted to ensure invariance to rotation, scaling, translation, partial illumination changes and projective transform in the image description. SIFT are computed during the features extraction phase, through a parallel task respect to \emph{RAG} creation.

\subsection{Graph construction}

\begin{figure}[ht!]
\centering
\subfloat[]{\includegraphics[scale=0.4]{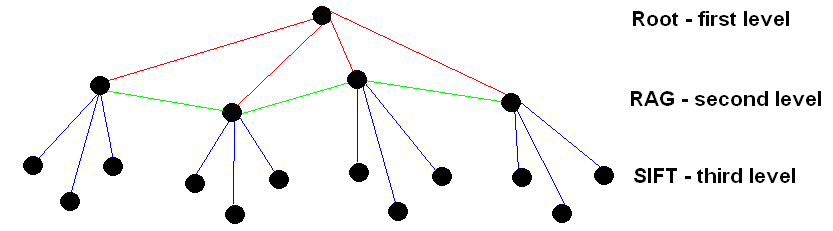}}\qquad
\subfloat[]{\includegraphics[scale=0.4]{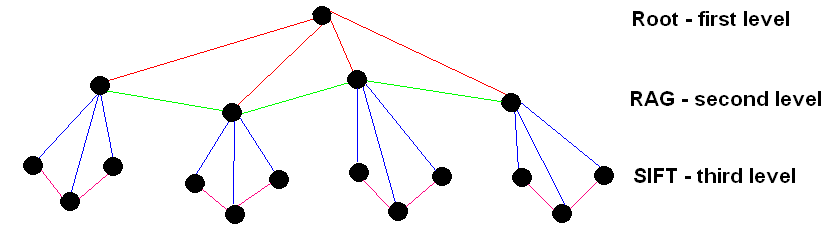}}\\
\caption{Region based (a) and Region graph based (b)
configurations.} \label{fig:GR}
\end{figure}

\textbf{ARSRG} building process consists in creation of three levels :
\begin{enumerate}
  \item \textbf{Root node}. The node located at the first level of graph structure and represents the image.
It is connected to all nodes at next level. 
\item \textbf{Region Adjacency Graph (RAG) nodes}. Adjacency relations among different image regions based on the segmentation result. Thus, adjacent image regions are represented by nodes connected at this level. 
  \item \textbf{Leaf nodes}. The set of SIFT features extracted from the image. Two type of connections are provided:
  \begin{enumerate}
    \item \emph{Region based}. A leaf node represents a SIFT keypoint obtained during
features extraction. Each leaf node-keypoint is associated to a
region based on its spatial coordinates in the image. At this level,
each node is connected with just one \emph{RAG} higher level node
(fig. \ref{fig:GR}(a)).
    \item \emph{Region graph based}. In addition to the previous configuration, leaf
nodes-keypoints belonging to the same region are connected by edges,
which encode spatial adjacency, based on a thresholding criteria (fig. \ref{fig:GR}(b)).
\end{enumerate}
\end{enumerate}

\section{Formal definitions}
\label{defs}


\textbf{ARSRG} structure is defined based on two leaf node configurations.

\newtheorem{mydef}{Definition}[section]

\begin{mydef}
\label{ARSRG1}

\textbf{ARSRG$_{1^{st}}$} (first leaf nodes configuration) $G$ is defined as a tuple
$G=(V_{regions},E_{regions},VF_{SIFT},E_{regions-SIFT})$, where

\begin{itemize}
  \item $V_{regions}$, the set of regions-nodes.
  \item $E_{regions} \subseteq V_{regions} \times V_{regions}$, the set of
undirected edges, where $e \in E_{regions}$ and $e=(v_{i},v_{j})$ is
an edge that connect nodes $v_{i},v_{j} \in V_{regions}$.
  \item $VF_{SIFT}$, the set of SIFT-nodes.
  \item $E_{regions-SIFT} \subseteq V_{regions} \times VF_{SIFT}$, the set of directed
  edges, where $e \in E_{regions-SIFT}$ and
$e=(v_{i},vf_{j})$ is an edge that connect source node $v_{i} \in
V_{regions}$ and destination node $vf_{j} \in VF_{SIFT}$.
\end{itemize}
\end{mydef}

\begin{mydef}
\label{ARSRG2} \textbf{ARSRG$_{2^{nd}}$} (second leaf nodes
configuration), $G$ is defined as a tuple
$G=(V_{regions},E_{regions},VF_{SIFT},E_{regions-SIFT},E_{SIFT})$,
where:

\begin{itemize}
\item $V_{regions}$, the set of regions-nodes.
\item $E_{regions} \subseteq V_{regions} \times V_{regions}$, the set
of undirected edges, where $e \in E_{regions}$ and $e=(v_{i},v_{j})$
is an edge that connect nodes $v_{i},v_{j} \in V_{regions}$
\item $VF_{SIFT}$, the set of SIFT-nodes.
\item $E_{regions-SIFT} \subseteq V_{regions} \times
VF_{SIFT}$, the set of directed edges, where $e \in
E_{regions-SIFT}$ and $e=(v_{i},vf_{j})$ is an edge that connect
source node $v_{i} \in V_{regions}$ and destination node $vf_{j} \in
VF_{SIFT}$.
\item $E_{SIFT} \subseteq VF_{SIFT} \times VF_{SIFT}$, the set of
undirected edges, where $e \in E_{SIFT}$ and $e=(vf_{i},vf_{j})$ is
an edge that connect nodes $vf_{i},vf_{j} \in V_{SIFT}$
\end{itemize}
\end{mydef}

\textbf{ARSRG} structures, first and second leaf node configuration,
are created based on definitions \ref{ARSRG1} and \ref{ARSRG2}. The
nodes belonging to sets $V_{regions}$ and $VF_{SIFT}$ are associated to features extracted from the image. Particularly:

\begin{mydef}
\label{RegFeat} $F_{regions}$ is a set of vectors attributes
associated to nodes in $V_{regions}$. An element, $f_{i} \in v_{i}$,
is associated to a node of \textbf{ARSRG} structure at second level.
It contains the region dimension (pixels).
\end{mydef}

\begin{mydef}
$F_{SIFT}$ is a set of vectors attributes associated to nodes in
$VF_{SIFT}$. An element, $f_{i} \in vf_{i}$, is associated to a node of \textbf{ARSRG} structure at third level. It contains a SIFT
descriptor. 
\end{mydef}

The association between features and nodes is performed through
assignment functions defined as follows:

\begin{mydef}
\label{AssRegFeat} The node-labeling function $L_{regions}$ assigns
a label to each node $v \in V_{regions}$ of \textbf{ARSRG} at the
second level. The node label is a feature attribute $d_{i}$
extracted from the image. The label value is the dimension of region
(pixels number). The labeling procedure of a $v$ node occurs during
the process of \textbf{ARSRG} construction.
\end{mydef}

\begin{mydef}
The SIFT node-labeling function $L_{SIFT}$ assigns a label to each
node $vf \in VF_{SIFT}$ of \textbf{ARSRG} at third level. The node
label is a features vector $f_{i}$, keypoint, extracted from the
image. The labeling procedure of a $vf$ node checks the position of keypoint in the image compared to the region to which it
belongs. 
\end{mydef}

Also, the \emph{RAG nodes} $\in V_{regions}$ are doubly linked in
horizontal order, between them, and vertical order, with nodes $\in
VF_{SIFT}$. Edges $\in E_{regions}$ are all undirected from left to
right. While, edges $\in E_{regions-SIFT}$ are all directed from top
to bottom. The \emph{Root node} maintains list of edges outgoing to \emph{RAG nodes}. Also, each \emph{RAG node} maintains three linked lists of edges: one for outgoing to \emph{RAG nodes}, one for outgoing \emph{leaf nodes} and one for ingoing to \emph{Root node}. Finally, each \emph{leaf node} maintains
two linked lists of edges: one for ingoing from \emph{RAG nodes} and
one for outgoing \emph{leaf nodes}. The edges in each list are
ordered based on distances between end nodes: shorter edges come
first. These lists of edges have direct geometrical meanings: each
node is connected to another node in one direction: left, right,
top, and bottom. 

A very important aspect concerns the organization of the third level of the \textbf{ARSRG} structure. To this end, SIFT Nearest-Neighbor
Graph ($SNNG$) is introduced.

\begin{mydef}
\label{Snng} A $SNNG=(VF_{SIFT},E_{SIFT})$ is defined as

\begin{itemize}
\item $VF_{SIFT}$: the set of nodes associated to SIFT
keypoints
\item $E_{SIFT}$: the set of edges, where for each $v_{i} \in VF_{SIFT}$, an edge
$(v_{i},v_{ip})$ if and only if $dist(v_{i},v_{ip})< \tau$ exists.
$dist(v_{i},v_{ip})$ is Euclidean distance applied to $x$ and $y$
position of keypoints in the image, $\tau$ is a threshold value and
$p$ stems from $1$ to $k$, $k$ being the size of $VF_{SIFT}$.
\end{itemize}
\end{mydef}

This notation is very useful during the matching phase. Indeed, each
$SNNG$ indicates the set of SIFT features belonging to image region,
with reference to definition \ref{ARSRG2}, and
represents SIFT features organized from local and spatial point of
view. A different version of $SNNG$ is called complete SIFT
Nearest-Neighbor Graph ($SNNGc$).

\begin{mydef}
\label{cSnng} A $SNNGc=(VF_{SIFT},E_{SIFT})$ is defined as

\begin{itemize}
\item $VF_{SIFT}$: the set of nodes associated to SIFT
keypoints
\item $E_{SIFT}$: the set of edges, where for each $v_{i} \in VF_{SIFT}$, an edge
$(v_{i},v_{ip})$ if and only if $dist(v_{i},v_{ip})< \tau$ exists.
$dist(v_{i},v_{ip})$ is Euclidean distance applied to $x$ and $y$
position of keypoints in the image, $\tau$ is a threshold value and
$p$ stems from $1$ to $k$, $k$ being the size of $VF_{SIFT}$. In
this case, $\tau$ is greater than the maximal distance between
keypoints.
\end{itemize}
\end{mydef}


Another important aspect concerns the difference between vertical
and horizontal relationships among nodes in the \textbf{ARSRG}
structure. Below these relations, edges, are defined.

\begin{mydef}
\label{relhor1} A region horizontal edge $e$, $e \in E_{regions}$,
is an undirected edge $e=(v_{i},v_{j})$ that connects nodes
$v_{i},v_{j} \in V_{regions}$.
\end{mydef}

\begin{mydef}
\label{relhor2} A SIFT horizontal edge $e$, $e \in E_{SIFT}$, is an
undirected edge $e=(vf_{i},vf_{j})$ that connects nodes
$vf_{i},vf_{j} \in V_{SIFT}$.
\end{mydef}

\begin{mydef}
\label{relver} A vertical edge $e$, $e \in E_{regions-SIFT}$, is an
directed edge $e=(v_{i},vf_{j})$ that connects nodes $v_{i} \in
V_{regions}$ and $vf_{j} \in VF_{SIFT}$ from source node $v_{i}$ to
destination node $vf_{j}$.
\end{mydef}

As can be noted horizontal edges connect nodes of the same level. While, vertical
edges connect nodes of different levels (second-third). Finally,
these relations are represented through adjacency matrices defined
below.

\begin{mydef}
The binary regions adjacency matrix $S_{regions}$ describes the
spatial relations among RAG nodes. An element $s_{ij}$ defines an
edge, $e=(v_{i},v_{j})$, connecting nodes $v_{i},v_{j} \in
V_{regions}$. Hence, an element $s_{ij} \in S_{regions}$ is set to
$1$ if node $v_{i}$ is connected to node $v_{j}$, $0$ otherwise.
\end{mydef}

\begin{mydef}
The binary SIFT adjacency matrix $S_{SIFT}$ describes the spatial
relations among leaf nodes. An element $s_{ij}$ defines an edge,
$e=(vf_{i},vf_{j})$, connecting nodes $vf_{i},vf_{j} \in VF_{SIFT}$.
Hence, an element $s_{ij} \in S_{SIFT}$ is set to $1$ if node
$vf_{i}$ is connected to node $vf_{j}$, $0$ otherwise.
\end{mydef}





\begin{figure}[H]
\centering
\subfloat[]{\includegraphics[width=0.45\textwidth]{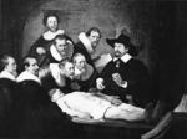}
 \label{fig:Rank1}}\qquad
\subfloat[]{\includegraphics[width=0.45\textwidth]{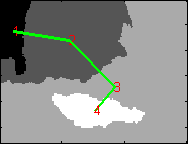}
 \label{fig:Rank2}}\\
\subfloat[]{\includegraphics[width=0.45\textwidth]{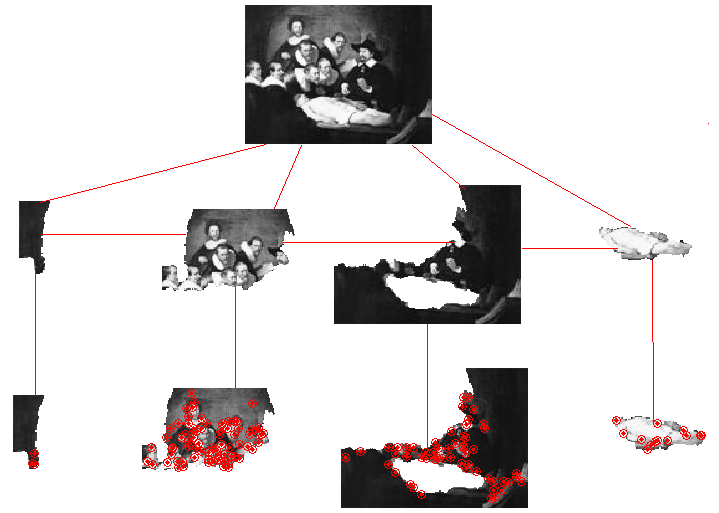}
 \label{fig:Rank3}}%
 \subfloat[]{\includegraphics[width=0.45\textwidth]{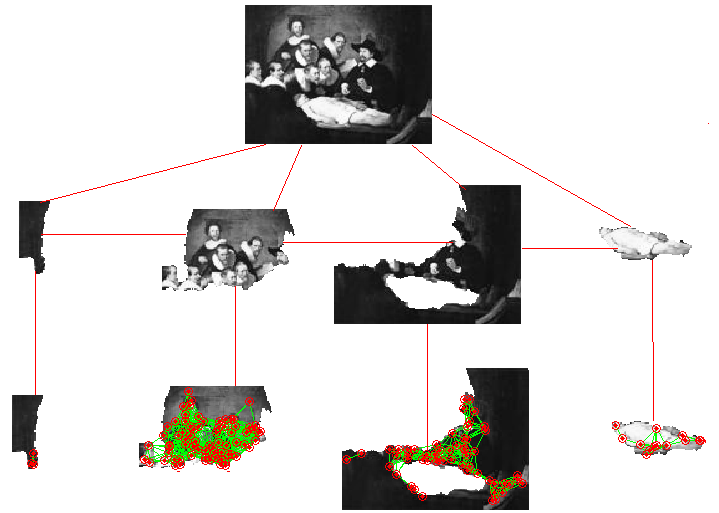}
 \label{fig:Rank4}}%
\caption{(a) Original image; (b) \emph{RAG} composed of $4$ regions;
(c) Region based leaf node configuration; (d) Region graph based
leaf node configuration. Red point in figures (c) and (d) represent
SIFT keypoints belonging to regions. While green lines in the figure
(d) represent the edges of graph based leaf node configuration.}
\label{ARSRGcon}
\end{figure}

Figures \ref{ARSRGcon} show the two different \textbf{ARSRG}
structures on a sample image.


\section{Properties}
\label{props}

In this section, \textbf{ARSRG} structure properties
arising from features extraction and graph construction steps are highlighted.\\

\textbf{Region features and structural information}. The main goal
of the \textbf{ARSRG} structure is to connect regional features and
structural information. First step concerns image segmentation in
order to extract ROIs. This is a step towards the extraction of
semantic information from a scene. Once the image has been
segmented, the \emph{RAG} structure is created. This features representation highlights individual regions and spatial relations existing between them.\\

\textbf{Horizontal and vertical relations}. \textbf{ARSRG} structure
presents two types of relations (edges) between image features:
horizontal and vertical. Vertical edges define image topological
structure, while horizontal edges define spatial constraints about nodes (regions) features. Horizontal relations (definitions
\ref{relhor1} and \ref{relhor2}) concern ROIs and SIFT features located at the second level of the structure. The general goal is to provide information of spatial closeness, define spatial constraints on the node attributes, characterize features map of specific resolution level (detail) on a
defined image and can be differentiated according
to the computational complexity and the occurrence frequency. Their order is in the range
$\{1,\dots,n\}$, where $n$ is the number of features specified through
the relations. In a
different way, vertical relations (definition \ref{relver}) concern
connections between individual regions and their features. The
vertical directed edges connect nodes among second and third levels
of \textbf{ARSRG} (\emph{RAG} nodes to \emph{leaf} nodes) and provide a parent-child relationship. In this context, the role of \textbf{ARSRG} structure is to create a bridge between the defined relations. This aspect leads to some advantages, i.e. the possibility to explore the structure both to in breadth and depth during the matching process.\\

\textbf{Region features invariant to point of view, illumination and
scale}. Building local invariant region descriptors is a
hot topic of research with a set of applications such as object
recognition, matching and reconstruction. Over the last years, great
success has been achieved in designing descriptors invariant to
certain types of geometric and photometric transformations. Local
Invariant Features Extraction (LIFE) methods work in order to
extract stable descriptors starting from a particular set of
characteristic regions of the image. LIFE methods were chosen, for
region representation, in order to provide invariance to certain
conditions. These local representations, created by using
information extracted from each region, are robust to certain image
deformations such as illumination and viewpoint changing.
\textbf{ARSRG} structure includes SIFT features, identified in
\cite{MikolajczykSchmid} as the most stable representations between
different LIFE methods.\\

\textbf{Advantages due to detailed information located on different level}. The detailed image description, provided by
the \textbf{ARSRG} structure, represents an advantage during the
comparison phase. In hierarchical way, the matching procedure
explores global, local and structural information, within
\textbf{ARSRG}. First step involves a filtering procedure for
regions based on size. Small regions, containing poor information,
are removed. Subsequently, the matching procedure goes to next level
of the \textbf{ARSRG} structure analyzing features of single regions
to obtain a stronger match. The goal is to solve the mapping on
multiple SNNGs (definition \ref{Snng}) of the \textbf{ARSRG}s. In
essence, this criterion identifies partial matches among SNNGs belonging to \textbf{ARSRG}s. During the
procedure, different combinations of graph SNNGs are identified and
a hierarchy of the matching process is constructed. In this way, the
overall complexity is reduced, which is expected to show
considerable
advantage especially for large \textbf{ARSRG}s.\\


\textbf{Advantages due to match region-by-region}. Region-Based
Image Retrieval (RBIR) \cite{LiuZhangLuMa} systems work with the
goal of extracting and defining similarity between two images based
on regional features. It has been demonstrated that users focus
their attention on specific regions rather than the entire image.
Region based image representation has proven to be more close
to human perception. In this context, in order to compare
\textbf{ARSRG} structures, a region matching scheme based on
appearance similarities of image segmentation results can be adopted.
Region matching algorithm exploits the regions provided by
segmentation and compares the features associated to them. The pairwise region similarities are computed from a set of SIFT features
belonging to regions. The matching procedure is asymmetric. The
input image is segmented into regions and its groups of SIFT
keypoints can be matched within consistent portion of the other
image. In this way, segmentation result is used to create regions of
candidate keypoints, avoiding incompatible regions for two images of
the same scene.\\ 

\textbf{False matches removal}. One of the main issues of LIFE
methods concerns the removal of false matches. It has been shown
that LIFE methods produce a number of false matches, during the
comparison phase, that significantly affect accuracy. The main reason concerns the lack of correspondence among image features (for example due to partial background occlusion of the scene). Standard similarity measures, based on the features descriptor, are widely used, even if they rely only on region appearance. In some cases, it cannot be sufficiently discriminating to ensure correct matches. This problem is more relevant in the presence of low or homogeneous textures, and leads to a lot of false matches. The application of the \textbf{ARSRG} structure provides a solution for this problem. In order to reduce false matches, small \textbf{ARSRG} regions-nodes, and associated SIFT descriptors, are removed. Indeed, small regions and their associated features are not very informative both in image description and matching. Ratio test \cite{Lowe} or graph matching \cite{SanromaManchoSerratosa} can be applied to perform comparison between remaining regions. This filtering procedure has a strong impact on experiments, resulting in a relevant accuracy improvement.

\section{Experimental results}
\label{res}

This section provides experimental results arising from different application fields. Particularly:

\begin{enumerate}
    \item Graph matching \cite{manzo2013attributed}. \textbf{ARSRG} is adopted to address the art painting retrieval problem. A graph matching algorithm is adopted to measure \textbf{ARSRG}s similarities exploiting local information and topological relations.
    \item Graph embedding \cite{manzo2014novel}. \textbf{ARSRG} is adopted to effectively tackle the object recognition problem. A framework to embed graph structures into vector space is built;
    \item Bag of Graph Words \cite{manzo2019bag}.  \textbf{ARSRG} is adopted to address the image classification problem. A digital image is described as a vector in terms of a frequency histogram of \textbf{ARSRG}s. 
    \item Kernel Graph Embedding \cite{manzo2019kgearsrg}. \textbf{ARSRG} is adopted to effectively tackle the imbalanced classification problem. A digital image is described through a vector-based representation called Kernel Graph Embedding on Attributed Relational Scale-Invariant Feature Transform-based Regions Graph (\textbf{KGEARSRG}).
\end{enumerate}

\subsection{Graph matching}
\label{gm}

\begin{figure}[!ht]
\centering
\subfloat[]{\includegraphics[width=0.45\textwidth]{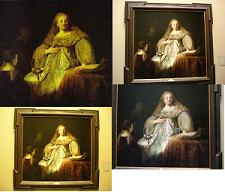}}\qquad
\subfloat[]{\includegraphics[width=0.45\textwidth]{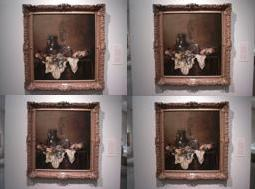}}\\
\caption{Some examples of art painting images.} \label{datasets:GR}
\end{figure}

In this section the results related to the work proposed in \cite{manzo2013attributed} are analyzed. \textbf{ARSRG} has been tested on three datasets and compared with LIFE methods, graph matching algorithms and a CBIR system. The first dataset, described in \cite{Haladova}, is
composed by two sets of images obtained from Olga's
gallery\footnote{http://www.abcgallery.com/index.html} and Travel
Webshots\footnote{http://travel.webshots.com}. The second dataset,
described in\cite{EtezadiAmoliChangHewlett}, is composed by painting
photos taken from the Cantor Arts Center\footnote{http://museum.stanford.edu/}.
The third dataset, described in \cite{BorisEffrosyniMarcinRufKokiopoulouDetyniecki}, is
composed by $1002$ images. Figure \ref{datasets:GR} shows some
examples.

\subsubsection{Discussion}

A first evaluation is performed for dataset used in \cite{Haladova}
and through comparisons with LIFE methods. Results are reported
in terms of Mean Reciprocal Rank (MRR). Table \ref{tab:MRRall} shows that \textbf{ARSRG} based approach provides best performance. As in \cite{Haladova,Lowe}, a tuning procedure is applied to $\rho$ parameter that controls tolerance of false matches both in graph matching and ratio test. In particular, $\rho$ values of $0.6$ and $0.7$ are used in \cite{Haladova} and values greater than $0.8$ are rejected as in \cite{Lowe}. $\rho$ values of $0.7$ and $0.8$ give optimal results for \textbf{ARSRG} matching. Graph based image representation clearly captures the topological relationships among features and acts as a filter over the complete set of SIFT features extracted from the image. Indeed, the comparison was performed among descriptors belonging to regions instead of entire image as proposed in standard approaches. In this way, many false matches are discarded and effectiveness is greatly improved.

\begin{table}[!ht]
\centering \caption{Quantitative comparison using $MRR$ measure
among SIFT\cite{Lowe}, SURF\cite{BayTuytelaarsGool},
ORB\cite{RubleeRabaudKonoligeBradski}, FREAK\cite{Alahi2012},
BRIEF\cite{CalonderLepetitStrechaFua} and \textbf{ARSRG} matching on dataset
in\cite{Haladova}.}
\begin{tabular}{|l|l|l|l|l|l|l|l|l|}
  \hline
  $\rho$ & \emph{SIFT} & \emph{SURF} & \emph{ORB} & \emph{FREAK} & \emph{BRIEF} & \textbf{ARSRG}$_{1^{st}}$ & \textbf{ARSRG}$_{2^{nd}}$\\ \hline
  $0.6$ & 0.7485 & 0.8400 & 0.6500 & 0.3558 & 0.4300 & 0.6700 & 0.6750\\ \hline
  $0.7$ & 0.7051 & 0.6800 & 0.6116 & 0.3360 & 0.3995 & 0.7133 & 0.7500\\ \hline
  $0.8$ & 0.6963 & 0.5997 & 0.5651 & 0.2645 & 0.4227 & 0.6115 & 0.8000\\ \hline
\end{tabular}
\label{tab:MRRall}
\end{table}

A second test has been performed on the dataset adopted in \cite{EtezadiAmoliChangHewlett}, computing performance in terms of Precision and Recall. Values of $\rho$ parameter are the
same as in the previous test. Table \ref{tab:Recall} shows that SIFT based approach performs better in terms of Recall. In case of $\rho$ equal to $0.8$, \textbf{ARSRG} matching yields comparable results.
In contrast, Table \ref{tab:Precision} shows that \textbf{ARSRG} matching, clearly outperforming the other approaches in terms of Precision,
proves to be very effective for image retrieval problem. The best
results by \textbf{ARSRG} matching algorithm for Precision are provided with
$\rho$ equal to $0.6$, $0.7$ and $0.8$. These results are due to the use of image structural representation. Indeed, \textbf{ARSRG} nodes,
representing different image regions, provide a partitioning rule
applied on entire set of SIFT. In this way, the subsets obtained are considered separately during matching step. This strategy removes most of false matches that normally belongs to accepted matches. As a consequence, several images are discarded as candidates for final ranking.

\begin{table}[!ht]
\centering \caption{Quantitative comparison, using Recall measure,
among SIFT\cite{Lowe}, SURF\cite{BayTuytelaarsGool},
ORB\cite{RubleeRabaudKonoligeBradski}, FREAK\cite{Alahi2012},
BRIEF\cite{CalonderLepetitStrechaFua} and \textbf{ARSRG} matching on dataset
in\cite{EtezadiAmoliChangHewlett}.}
\begin{tabular}{|l|l|l|l|l|l|l|l|}
\hline
  $\rho$ & \emph{SIFT} & \emph{SURF} & \emph{ORB} & \emph{FREAK} & \emph{BRIEF} & \textbf{ARSRG}$_{1^{st}}$ & \textbf{ARSRG}$_{2^{nd}}$\\ \hline
  $0.6$ & 1.0 & 0.8666 & 0.8000 & 0.7333 & 0.7666 & 0.7333 & 0.7333\\ \hline
  $0.7$ & 1.0 & 0.9000 & 0.8666 & 0.7333 & 0.8666 & 0.7666 & 0.7333\\ \hline
  $0.8$ & 1.0 & 1.0 & 1.0 & 0.8333 & 1.0000 & 0.8000 & 0.8000\\ \hline
\end{tabular}
\label{tab:Recall}
\end{table}

\begin{table}[!ht]
\centering \caption{Quantitative comparison using Precision
measure, among SIFT\cite{Lowe}, SURF\cite{BayTuytelaarsGool},
ORB\cite{RubleeRabaudKonoligeBradski}, FREAK\cite{Alahi2012},
BRIEF\cite{CalonderLepetitStrechaFua} and \textbf{ARSRG} matching on dataset
in\cite{EtezadiAmoliChangHewlett}.}
\begin{tabular}{|l|l|l|l|l|l|l|l|}
\hline
  $\rho$ & \emph{SIFT} & \emph{SURF} & \emph{ORB} & \emph{FREAK} & \emph{BRIEF} & \textbf{ARSRG}$_{1^{st}}$ & \textbf{ARSRG}$_{2^{nd}}$\\ \hline
  $0.6$ & 0.0674 & 0.0820 & 0.2051 & 0.05584& 0.10689 & 1.0 & 1.0\\ \hline
  $0.7$ & 0.0401 & 0.0441 & 0.0742 & 0.04671& 0.05664 & 1.0 & 1.0\\ \hline 
  $0.8$ & 0.0312 & 0.0338 & 0.0348 & 0.04072& 0.03452 & 1.0 & 1.0\\ \hline 
\end{tabular}
\label{tab:Precision}
\end{table}

Additional experiments concern comparisons with graph SIFT-based
matching algorithms. Experiments are performed on datasets presented in \cite{Haladova,BorisEffrosyniMarcinRufKokiopoulouDetyniecki} and are evaluated through MRR measure. Results are reported in tables \ref{tab:MRRallGM1} and \ref{tab:MRRallGM2} and show comparison with HGM \cite{LeeChoLee}, RRWGM \cite{LeeChoLee2}, TM \cite{DuchenneBachKweonPonce} algorithms. Also in this case, \textbf{ARSRG} leads to better results compared to those obtained by the other graph SIFT-based matching algorithms. Similarly in this case, the region matching approach, by providing local information about spatial distribution of the features, leads to false matches removal and hence improves final results.

\begin{table}[!ht]
\centering \caption{Quantitative comparison, using $MRR$ measure,
among HGM\cite{LeeChoLee}, RRWGM\cite{LeeChoLee2},
TM\cite{DuchenneBachKweonPonce} algorithms and \textbf{ARSRG} matching on
dataset in\cite{Haladova}.}
\begin{tabular}{|l|l|l|l|l|}
\hline
  \emph{HGM} & \emph{RRWGM} & \emph{TM} & \textbf{ARSRG}$_{1^{st}}$ & \textbf{ARSRG}$_{2^{nd}}$\\ \hline
  0.2600 & 0.1322 & 0.1348 & 0.6115 & 1.0\\ \hline 
\end{tabular}
\label{tab:MRRallGM1}
\end{table}

\begin{table}[!ht]
\centering \caption{Quantitative comparison, using $MRR$ measure,
among HGM\cite{LeeChoLee}, RRWGM\cite{LeeChoLee2},
TM\cite{DuchenneBachKweonPonce} algorithms and \textbf{ARSRG} matching on
dataset in\cite{BorisEffrosyniMarcinRufKokiopoulouDetyniecki}.}
\begin{tabular}{|l|l|l|l|l|}
\hline
  \emph{HGM} & \emph{RRWGM} & \emph{TM} & \textbf{ARSRG}$_{1^{st}}$ & \textbf{ARSRG}$_{2^{nd}}$\\ \hline
  0.1000 & 0.0545 & 0.0545 & 0.20961 & 0.39803\\ \hline
\end{tabular}
\label{tab:MRRallGM2}
\end{table}

Final experiments concern  performance comparison with Lucene Image
Retrieval (LIRe) \cite{LuxChatzichristofis} system and some features available: \emph{MPEG7}\cite{MPEG7}, \emph{Tamura}\cite{Tamura}, \emph{CEDD}\cite{CEDD}. \emph{FCTH}\cite{FCTH}, \emph{ACC}\cite{CC}. Experiments are
performed on dataset presented in \cite{Haladova}, considering different features implemented in LIRe, and evaluated through MRR
measure. Results are reported in table \ref{tab:MRRLire}. It is clear that LIRe system is not very suitable for art paint retrieval, due to its low performing features, which results in wrong discrimination of relevant and irrelevant images. Consequently, the achieved ranking contains inadequate results, with respect to user's request, which affects heavily its final performance. In contrast, results obtained by \textbf{ARSRG} demonstrates once more that is very effective for this application.

\begin{table}[!ht]
\centering
\caption{Quantitative comparison using $MRR$ measure,
among some features available in (LIRe)\cite{LuxChatzichristofis}
system and \textbf{ARSRG} matching on dataset in\cite{Haladova}.}

\begin{tabular}{|l|l|l|l|l|l|l|l|l|l|}
  \hline
  \emph{MPEG7} & \emph{Tamura} & \emph{CEDD} & \emph{FCTH} & \emph{ACC} & \textbf{ARSRG}$_{1^{st}}$ & \textbf{ARSRG}$_{2^{nd}}$\\ \hline
  0.2645 & 0.1885 & 0.2329 & 0.1924 & 0.1879 & 0.7133 & 0.7500\\ \hline
\end{tabular}
\label{tab:MRRLire}
\end{table}

\subsection{Graph embedding}
\label{ExpGe} 
In this section the results related to the work proposed in \cite{manzo2014novel} are analyzed. \textbf{ARSRG} on three popular datasets, that differs in size, design, and topic, about well-known object recognition field is tested. Precisely, the following databases are employed:

\begin{enumerate}
  \item The Columbia Image Database Library (COIL-$100$) \cite{nayar1996columbia}, which consists of $100$ objects. Each object is represented by $72$ colored images that show it under different rotation point of view. The objects have been located on a black background.
  \item The Amsterdam Library Of Images (ALOI) \cite{geusebroek2005amsterdam} is a color image collection of $1000$ small objects. In contrast to COIL-$100$, where the objects are cropped to fill the full image, in ALOI the images contain the background and the objects in their original size. The objects have been located on a black background.
\item The ETH-$80$ \cite{leibe2003analyzing}, which contains $80$ objects from $8$
categories and each object is represented by $41$ different views, thus obtaining a total of $3280$ images. The objects have been located on a uniform background.
\end{enumerate}

\begin{figure} [!ht]
\centering
\subfloat[]{\includegraphics[scale=0.3]{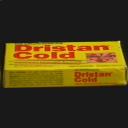}}\qquad
\subfloat[]{\includegraphics[scale=0.3]{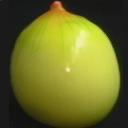}}\qquad
\subfloat[]{\includegraphics[scale=0.26]{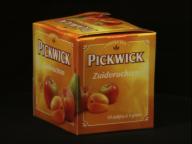}}\qquad
\subfloat[]{\includegraphics[scale=0.26]{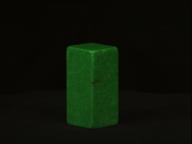}}\qquad
\subfloat[]{\includegraphics[scale=0.3]{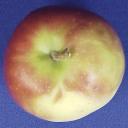}}\qquad
\subfloat[]{\includegraphics[scale=0.3]{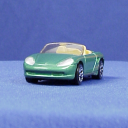}}\qquad
\caption{Example images from the COIL-$100$ dataset (a,b), ALOI dataset (c,d) and ETH-$80$ dataset (e,f).} \label{fig:ds}
\end{figure}

\subsubsection{Discussion}
\label{discussion}

Table \ref{ETH-80ge} summarizes the accuracy results of the proposed
framework on ETH-$80$ database. In order to perform a direct
comparison with the methods employed in~\cite{morales2014new}, the
same setup is adopted. Precisely, the same $6$ categories
(\emph{apples}, \emph{cars}, \emph{cows}, \emph{cups},
\emph{horses}, and \emph{tomatoes}) are adopted. For each category $4$ objects
are taken and for each object $10$ different views are considered thus obtaining a
total of $240$ images. From the remaining images, $60$ per category
($15$ views per object) are used as testing examples. The results are achieved by baseline Logistic Label Propagation ($LLP$) \cite{kobayashi2012logistic} + Bag of Words (BoW) \cite{lazebnik2006beyond}), and those obtained
in~\cite{morales2014new} by employing the approaches proposed in
\cite{gago2010full} (gdFil), in \cite{jia2011efficient}
(APGM), and in \cite{acosta2012frequent} (VEAM). As can be seen in table \ref{ETH-80ge}, \textbf{ARSRG} embedding, adopting $LLP$ classifier, outperforms the results obtained by the other approaches. These results confirm that \textbf{ARSRG} embedding correctly deals with object view changes.

\begin{table}[!ht]
\caption{Recognition accuracy on the ETH-$80$ database.}
\label{ETH-80ge}\centering \tabcolsep=0.11cm
\begin{tabular}{|c|c|}
\hline
\textbf{Method} & \textbf{Accuracy}\\
\hline
\textbf{LLP}+\textbf{ARSRG}emb & \textbf{89.26}\%\\
\hline
\textbf{LLP}+BoW & 58.83\%\\
\hline
gdFil& 47.59\%\\
\hline
APGM & 84.39\%\\
\hline
VEAM & 82.68\%\\
\hline
\end{tabular}
\end{table}


Table \ref{COILge} summarizes the results achieved by
$LLP+$\textbf{ARSRG} on COIL-$100$ database. In order to
perform a direct comparison with the methods employed
in~\cite{morales2014new,morales2013simple}, the same setup is
adopted. Precisely, $25$ objects are randomly selected and the 11\% of the images as training set and the remaining
ones as testing set are selected. The results are achieved by baseline Logistic Label Propagation ($LLP$) \cite{kobayashi2012logistic} + Bag of Words (BoW) \cite{lazebnik2006beyond}), and those obtained
in~\cite{morales2014new,morales2013simple} by
employing their approach (VFSR) and the approaches proposed in \cite{gago2010full}
(gdFil), in \cite{jia2011efficient} (APGM), in
\cite{acosta2012frequent} (VEAM), in \cite{wang2006tensor}
(DTROD-AdaBoost), in \cite{maree2005decision}
(RSW+Boosting), in \cite{morioka2008learning}
(Sequential Patterns), and in \cite{obdrzalek2002object}
(LAF). The results are presented in terms of accuracy and the best performance is highlighted in bold face. As can be notice \textbf{ARSRG} embedding confirms its qualities also employing this database. Indeed \textbf{ARSRG} embedding obtained the best overall accuracy.

\begin{table}[!ht]
\caption{Recognition accuracy on the COIL-$100$ database.}
\label{COILge}\centering \tabcolsep=0.11cm
\begin{tabular}{|c|c|}
\hline
\textbf{Method} & \textbf{Accuracy}\\
\hline
$LLP$+\textbf{ARSRG}$emb$ & \textbf{99.55}\%\\
\hline
LLP+BoW & 51.71\%\\
\hline
gdFil & 32.61\%\\
\hline
VFSR& 91.60\%\\
\hline
APGM & 99.11\%\\
\hline
VEAM & 99.44\%\\
\hline
DTROD-AdaBoost  & 84.50\%\\
\hline
RSW+Boosting & 89.20\%\\
\hline
Sequential Patterns  & 89.80\%\\
\hline
LAF & 99.40\%\\
\hline
\end{tabular}
\end{table}




Table \ref{ALOIge} summarizes the accuracy results obtained on the ALOI database. In order to perform a direct comparison with the methods employed
in~\cite{uray2007incremental}, the same setup is adopted; precisely,
only the first $100$ objects are employed. Color images have been
converted to gray level and second image of each class was adopted
for training and the remaining for testing. Two images of each class are considered,
having a total of $200$ images. Subsequently, at each iteration for each class one
additional training image is attached. In table
\ref{ALOIge} only the results by considering batch of $400$ images are shown since the intermediate results did not provide great
differences. The results achieved by baseline
Logistic Label Propagation ($LLP$) \cite{kobayashi2012logistic} + Bag of Words (BoW) \cite{lazebnik2006beyond}), and those obtained
in~\cite{uray2007incremental} by employing some variants of Linear
Discriminant Analysis (ILDAaPCA, batchLDA, ILDAonK, and ILDAonL) are reported. These results show that $LLP+$\textbf{ARSRG}$emb$ is able to obtain good performance with a small amount of training set and that it is little affected by overfitting problems.

\begin{table}[!ht]
\caption{Recognition accuracy on the ALOI database.}
\label{ALOIge}\centering \tabcolsep=0.11cm
\tiny{
\begin{tabular}{|c|c|c|c|c|c|c|c|c|c|c|}
\hline
\textbf{Method} & \textbf{200} & \textbf{400} & \textbf{800} & \textbf{1200} & \textbf{1600} & \textbf{2000} & \textbf{2400} & \textbf{2800} & \textbf{3200} & \textbf{3600}\\
\hline
\multirow{1}{*}{$LLP+$\textbf{ARSRG}$emb$} & \textbf{86.00}\% & \textbf{90.00}\%  & \textbf{93.00}\%  & \textbf{96.00}\% & \textbf{95.62}\% & \textbf{96.00}\% & \textbf{88.00}\% & \textbf{81.89}\% & \textbf{79.17}\% & \textbf{79.78}\%\\
\hline
\multirow{1}{*}{\textbf{LLP}+\textbf{BoW}} & 49.60\% & 55.00\% & 50.42\% & 50.13\% & 49.81\% & 48.88\% & 49.52\% & 49.65\% & 48.96\% & 49.10\%\\
 \hline
\multirow{1}{*}{batchLDA}  & 51.00\% & 52.00\% & 62.00\% & 62.00\% & 70.00\% & 71.00\% & 74.00\% & 75.00\% & 75.00\% & 77.00\%\\
 \hline
\multirow{1}{*}{ILDAaPCA}  & 51.00\% & 42.00\% & 53.00\% & 48.00\% & 45.00\% & 50.00\% & 51.00\% & 49.00\% & 49.00\% & 50.00\%\\
 \hline
\multirow{1}{*}{ILDAonK}  & 42.00\% & 45.00\% & 53.00\% & 48.00\% & 45.00\% & 51.00\% & 51.00\% & 49.00\% & 49.00\% & 50.00\%\\
 \hline
\multirow{1}{*}{ILDAonL}  & 51.00\% & 52.00\% & 61.00\% & 61.00\% & 65.00\% & 69.00\% & 71.00\% & 70.00\% & 71.00\% & 72.00\%\\
 \hline
\end{tabular}
}
\end{table}


Moreover, results confirm that capturing local information preserving the spatial relationships between them can strongly improve the performance in the object recognition field. It is important to highlight that, thanks to graph embedding paradigm, the main computational overhead concerns only the
extraction of graph-based representation in the training stage,
while the classification can be performed very quickly. 




\subsection{Bag of \textbf{ARSRG} Words}
\label{BoGW}
In this section the results related to the work proposed in \cite{manzo2019bag}, named Bag of \textbf{ARSRG} Words (\textbf{BoAW}), are analyzed.
\textbf{BoAW} is tested on datasets, ALOI, COIL-100 and ETH-80, described in section \ref{ExpGe} and in addition on dataset described below: \begin{enumerate}
            \item Caltech 101 \cite{fei2007learning}. It is an objects image collection belonging to 101 categories, with about 40 to 800~images per category. Most categories have about 50 images. 

\end{enumerate}

\begin{figure}[!ht]
\centering
\subfloat[]{\includegraphics[width=0.2\textwidth]{./19_r135.jpg}
    }\qquad
\subfloat[]{\includegraphics[width=0.2\textwidth]{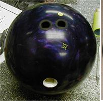}
    }\\
\subfloat[]{\includegraphics[width=0.2\textwidth]{./obj2__45.jpg}
    }%
    \subfloat[]{\includegraphics[width=0.2\textwidth]{./apple.jpg}
    }%
\caption{Dataset images: (\textbf{a}) ALOI, (\textbf{b}) Caltech 101, (\textbf{c}) COIL-100, (\textbf{d}) ETH-80.} \label{ds}
\end{figure}

ALOI, COIL-100 and ETH-80 datasets are represented on a simple background then the classification is less difficult than the dataset Caltech 101 where images {have a not uniform background}. Figure \ref{ds} shows some examples of datasets.

\subsubsection{Discussion}

The classification stage is managed with $LLP$ \cite{kobayashi2012logistic}. Tests are performed using a one-versus-all (OvA) paradigm for $30$ executions. A shuffling operation is applied to ensure the training and test set are always different. Images are scaled to $150 \times 150$ pixels size, to avoid performance degradation. Table \ref{ALOI} reports experiments performed on the ALOI dataset. Results are listed in order of average accuracy and the approach that provided the best performance is highlighted. In order to perform a correct comparison the same settings reported in~\cite{uray2007incremental} and related to table \ref{ALOIge} are adopted. The results in table \ref{ALOI} achieved by Bag of Visual Words (BoVW) \cite{lazebnik2006beyond} and those obtained in ~\cite{uray2007incremental} using some variants of linear discriminant analysis (ILDAaPCA, batchLDA, ILDAonK and ILDAonL) and in \cite{manzo2014novel} (\textbf{ARSRG}$emb$) are shown.

\begin{table}[!ht]
\caption{Results on the ALOI dataset.}
\label{ALOI}
\centering \tabcolsep=0.11cm
\tiny{
\begin{tabular}{|c|c|c|c|c|c|c|c|c|c|c|}
\hline
\textbf{Method} & \textbf{200} & \textbf{400} & \textbf{800} & \textbf{1200} & \textbf{1600} & \textbf{2000} & \textbf{2400} & \textbf{2800} & \textbf{3200} & \textbf{3600}\\
\hline
\multirow{1}{*}{\textbf{BoAW} } & \textbf{98.29
}\% & \textbf{92.83}\% & \textbf{98.80}\% & \textbf{96.80}\% & \textbf{96.76}\% & \textbf{98.15}\% & \textbf{89.52}\% & \textbf{82.65}\% & \textbf{79.96}\% & \textbf{79.88}\%\\
\hline
\multirow{1}{*}{\textbf{ARSRG}$emb$} & 86.00\% & 90.00\%  & 93.00\%  & 96.00\% & 95.62\% & 96.00\% & 88.00\% & 81.89\% & 79.17\% & 79.78\%\\
\hline
\multirow{1}{*}{BoVW} & 49.60\% & 55.00\% & 50.42\% & 50.13\% & 49.81\% & 48.88\% & 49.52\% & 49.65\% & 48.96\% & 49.10\%\\
\hline
\multirow{1}{*}{batchLDA}  & 51.00\% & 52.00\% & 62.00\% & 62.00\% & 70.00\% & 71.00\% & 74.00\% & 75.00\% & 75.00\% & 77.00\%\\
 \hline
\multirow{1}{*}{ILDAaPCA}  & 51.00\% & 42.00\% & 53.00\% & 48.00\% & 45.00\% & 50.00\% & 51.00\% & 49.00\% & 49.00\% & 50.00\%\\
\multirow{1}{*}{ILDAonK}  & 42.00\% & 45.00\% & 53.00\% & 48.00\% & 45.00\% & 51.00\% & 51.00\% & 49.00\% & 49.00\% & 50.00\%\\
 \hline
\multirow{1}{*}{ILDAonL}  & 51.00\% & 52.00\% & 61.00\% & 61.00\% & 65.00\% & 69.00\% & 71.00\% & 70.00\% & 71.00\% & 72.00\%\\
 \hline
\end{tabular}
}
\end{table}


As can be seen, \textbf{BoAW} is able to provide best performance for the object recognition task. Indeed, the combination of local and spatial information provides clear benefits in image representation and matching.

Table \ref{tabcaltech} summarizes the results obtained on the Caltech 101 dataset. Experimental results are performed comparing the \textbf{BoAW}  with BoVW based on pyramidal representation \cite{lazebnik2006beyond}. Experimental~comparisons are performed using the following image categories: bowling, cake, calculator, cannon, cd, chess-board, joy-stick, skateboard, spoon and umbrella. The best performances are obtained with a training set and the test set at 60\% and 40\% of the dataset respectively. Results are listed in form of average accuracy and the approach that provided the best performance is  highlighted.

\begin{table}[!ht]
\centering
\caption{Results on the Caltech 101 dataset.}
\begin{tabular}{|c|c|c|c|c|c|c|}
\hline
\textbf{Method}  & \textbf{Accuracy}\\
\hline
BoAW             &     74.00\%    \\
\hline
 BoVW               & \textbf{83.00\%}   \\
\hline
\end{tabular}
\label{tabcaltech}
\end{table}

As can be seen the performance differs when images are composed of non-uniform backgrounds. BoVW is more efficient and does not suffer this detail otherwise decisive for \textbf{BoAW} which incorporates structural information. This aspect considerably distorts image representation and consequently the classification phase. This problem could be solved with a segmentation phase, during the preprocessing, to remove the uninformative background or with a filtering application, thus going to work exclusively on the object to be represented. This loophole does not always work because removing the background is not easy. Table \ref{COIL} shows results on the COIL-100 dataset using the same setup of table \ref{COILge}. Therefore, results obtained by BoVW are shown and those obtained in~\cite{morales2014new,morales2013simple} by applying their solution (VFSR) and the approaches proposed in \cite{gago2010full}
(gdFil), in \cite{jia2011efficient} (APGM), in~\cite{acosta2012frequent} (VEAM), in \cite{wang2006tensor}
(DTROD-AdaBoost), in \cite{maree2005decision}
(RSW+Boosting), in \cite{morioka2008learning}
(Sequential~Patterns), in \cite{obdrzalek2002object}
(LAF) and in \cite{manzo2014novel} (\textbf{ARSRG}$emb$). Results are listed in form of average accuracy and the approach that provided the best performance is  highlighted. Also in this case \textbf{BoAW} confirms its qualities obtaining the best performance.
 

\begin{table}[!ht]
\caption{Results on the COIL-100 dataset.}
\label{COIL}
\centering \tabcolsep=0.11cm
\begin{tabular}{|c|c|}
\hline
\textbf{Method} & \textbf{Accuracy}\\
\hline
\textbf{BoAW}  & \textbf{99.77}\%\\
\hline
\textbf{ARSRG}$emb$ & 99.55\%\\
\hline
BoVW & 51.71\%\\
\hline
gdFil & 32.61\%\\
\hline
VFSR& 91.60\%\\
\hline
APGM & 99.11\%\\
\hline
VEAM & 99.44\%\\
DTROD-AdaBoost  & 84.50\%\\
\hline
RSW+Boosting & 89.20\%\\
\hline
Sequential Patterns  & 89.80\%\\
\hline
LAF & 99.40\%\\
\hline
\end{tabular}
\end{table}

Table \ref{ETH-80} shows results on the ETH-80 dataset using the setup related to table \ref{ETH-80ge}. Tests performed by BoVW and those achieved in \cite{morales2014new} by employing the solution proposed in \cite{manzo2014novel} (\textbf{ARSRG}$emb$), \cite{gago2010full} (gdFil), in \cite{jia2011efficient} (APGM), and in \cite{acosta2012frequent} (VEAM) are presented. Also in this case the results are listed highlighting the accuracy of the best approach. As can be seen \textbf{BoAW} provides better results than competitors also when view points changes occur.


\begin{table}[!ht]
\caption{Results on the ETH-80 dataset.}
\label{ETH-80}
\centering \tabcolsep=0.11cm
\begin{tabular}{|c|c|}
\hline
\textbf{Method} & \textbf{Accuracy}\\
\hline
\textbf{BoAW}  & \textbf{89.29}\%\\
\hline
\textbf{ARSRG}$emb$ & 89.26\%\\
\hline
BoW & 58.83\%\\
\hline
{gdFil}& 47.59\%\\
\hline
{APGM} & 84.39\%\\
\hline
{VEAM} & 82.68\%\\
\hline
\end{tabular}
\end{table}

\subsection{Kernel Graph embedding}

In this section the results related to the work proposed in \cite{manzo2019kgearsrg} are analyzed. The classification performance through Support Vector machine (SVM) and Asimmetric Kernel Scaling (AKS) \cite{maratea2011asymmetric} over the standard OvA paradigm on different low, medium and high imbalanced image classification problems is tested, with art painting classification application~\cite{CuljakMikusJezHadjic}. The datasets adopted are the same described in section \ref{gm}. Tables \ref{ds1} and \ref{ds2} show settings about the classification problems. To notice, last column includes the imbalance rate (IR) calculated through equation \ref{IR}.


 {\begin{equation}
\label{IR}
IR=\frac{\%maj}{\%min}    
\end{equation}}

IR is defined as the ratio between the percentage of images belonging to the majority class over the minority class.


\begin{table}[!ht]

\footnotesize
\caption{OvA configuration for the dataset in~\cite{Haladova}.}
\centering
\begin{tabular}{|c|c|c|c|c|c|}
 \hline
 \textbf{Problem} & \textbf{Classification Problem}   & \textbf{(\%min,\%maj)} & \textbf{IR}\\ 
 \hline
 1 & Artemisia vs. all    & (3.00,97.00) & 32.33\\
 \hline
 2 & Bathsheba vs. all    & (3.00,97.00) & 32.33\\
\hline
3 & Danae vs. all    & (12.00,88.00) & 7.33\\ 
\hline
4 &
Doctor\_Nicolaes vs. all    & (3.00,97.00) & 32.33\\
\hline
5 & HollyFamilly vs. all    & (2.00,98.00) & 49.00\\
\hline
6 & PortraitOfMariaTrip vs. all    & (3.00,97.00) &
32.33\\
\hline
 7 & PortraitOfSaskia vs. all    &
(1.00,99.00) & 99.00\\
\hline
 8 & RembrandtXXPortrai vs. all  & (2.00,98.00) & 49.00\\
 \hline
  9 & SaskiaAsFlora vs. all  & (3.00,97.00) & 32.33\\
  \hline
   10 & SelfportraitAsStPaul vs. all  & (8.00,92.00) & 11.50\\
   \hline
    11 & TheJewishBride vs. all  & (4.00,96.00) & 24.00\\
    \hline
     12 & TheNightWatch vs. all  & (9.00,91.00) & 10.11\\
     \hline
      13 & TheProphetJeremiah vs
all  & (7.00,93.00) & 13.28\\
\hline
 14 &
TheReturnOfTheProdigalSon vs. all     & (9.00,91.00) &
10.11\\
\hline
 15 & TheSyndicsoftheClothmakersGuild vs. all 
& (5.00,95.00) & 19.00\\
 \hline
 16 & Other vs. all  &
(26.00,74.00) & 2.84\\
\hline
\end{tabular}
\label{ds1}
\end{table}
\begin{table}[!ht]
\caption{The OvA configuration for the dataset in~\cite{manzo2013attributed}.}
\centering
\begin{tabular}{|c|c|c|c|c|c|}
 \hline
 \textbf{Problem} & \textbf{Classification Problem}   & \textbf{(\%min,\%maj)} & \textbf{IR}\\ 
\hline
1 & Class 4 vs. all   & (1.00,9.00) & 9.00\\
\hline
2 & Class 7 vs. all  & (1.00,9.00) & 9.00\\
\hline
3 & Class 8 vs. all   & (1.00,9.00) & 9.00\\
\hline
4 & Class 13 vs. all  & (1.00,9.00) & 9.00\\
\hline
5 & Class 15 vs. all  & (1.00,9.00) & 9.00\\
\hline
6 & Class 19 vs. all  & (1.00,9.00) & 9.00\\
\hline
7 & Class 21 vs. all  & (1.00,9.00) & 9.00\\
\hline
8 & Class 27 vs. all  & (1.00,9.00) & 9.00\\
\hline
9 & Class 30 vs. all  & (1.00,9.00) & 9.00\\
\hline
10 & Class 33 vs. all  & (1.00,9.00) & 9.00\\
\hline
\end{tabular}

\label{ds2}
\end{table}











\subsubsection{Discussion}

\begin{figure}[!ht]
    \centering
     \subfloat[ \label{fig:RPC1}]{%
       \includegraphics[width=0.35\textwidth]{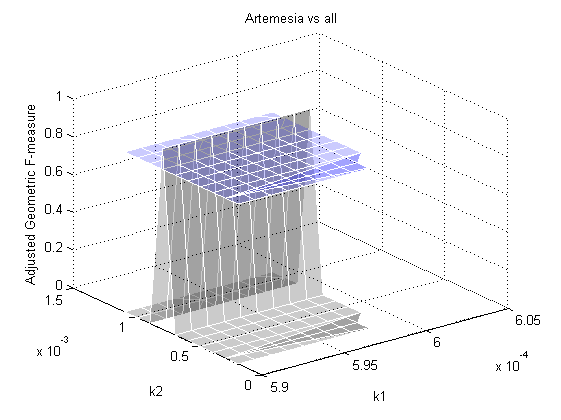}
     }
     \subfloat[ \label{fig:RPC2}]{%
       \includegraphics[width=0.35\textwidth]{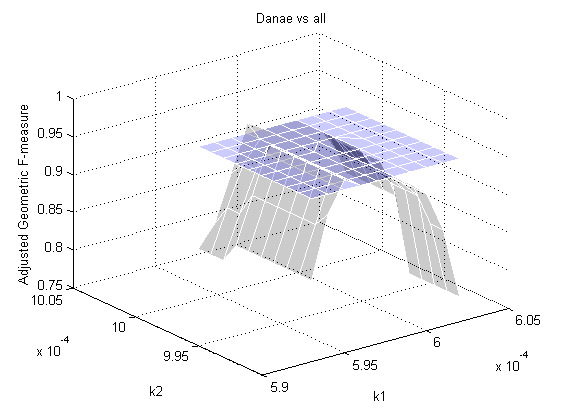}
    
     }
     
   \caption{Parameter choice 1. The $x$ and $y$ axes represent the
values of the parameters of the two methods, while on the $z$ axis is plotted
the AGF for two of the OvA configurations of the dataset in~\cite{Haladova}:
({\bf a}) Artemisia vs. all and ({\bf b}) Danae vs. all.  {The gray and blue surfaces represent, respectively, the results with the AKS and SVM classifiers.}} \label{parameters1}
   \end{figure}
\begin{figure}[!ht]
    \centering
     \subfloat[ \label{fig:RPC1}]{%
       \includegraphics[width=0.35\textwidth]{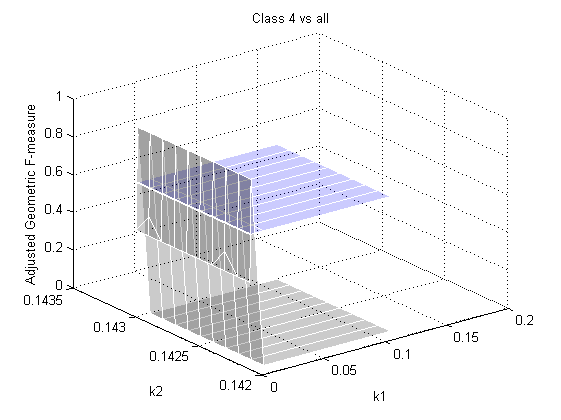}
     }
     \subfloat[ \label{fig:RPC2}]{%
       \includegraphics[width=0.35\textwidth]{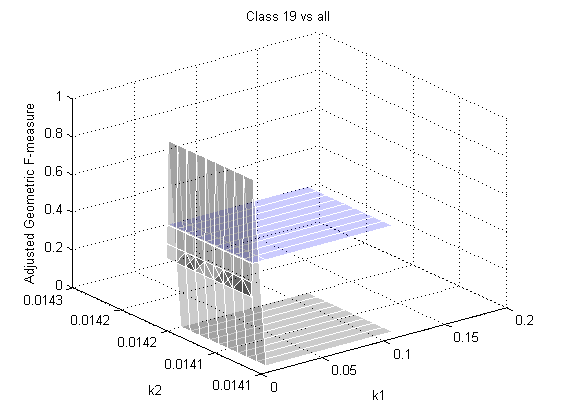}
    
     }
     
     \caption{Parameter choice 2. The $x$ and $y$ axes represent the
values of the parameters of the two methods, while on the $z$ axis is plotted the AGF for two of the OvA configurations on the dataset in~\cite{manzo2013attributed}: ({\bf a}) Class 4 vs. all and ({\bf b}) class 19 vs. all.  {The gray and blue surfaces represent, respectively, the results with the AKS and SVM classifiers.}}
\label{parameters2}
   \end{figure}

This section describes the comparison between AKS and standard SVM. The performance are described in terms of Adjusted F-measure \cite{maratea2013adjusted}. It can be seen in figure \ref{parameters1} that in order to reach noteworthy performance a fine tuning is needed and AKS consistently dominates standard SVM. Differently, in figure \ref{parameters2} performance presents only a single peak of exceedance with respect to SVM. Further tests have been performed in order to make a comparison with C4.5~\cite{Quinlan}, RIPPER~\cite{Cohen}, L2 Loss SVM \cite{BoserGuyonVapnik}, L2 Regularized Logistic Regression \cite{FanChangHsiehWangLin} and ripple-down rule learner (RDR)~\cite{WarnerJohnsonVamplew} for a complete set of OvA classification problems. The results of the two datasets are different due to imbalance rates. In the dataset in~\cite{Haladova}, configuration includes approximately low, medium and high rates. It is a great dataset for a robust testing phase because it covers full cases of
class imbalance problems. In the dataset in~\cite{manzo2013attributed}, imbalance rates are identical for all configurations. Results are reported in tables \ref{Res1}, for dataset in \cite{Haladova}, and \ref{Res2}, for dataset in \cite{manzo2013attributed}. It can be seen that performances are significantly higher than competitors. The improvement provided by AKS lies in the accuracy of the classification of patterns belonging to the minority class, positive, which, during the relevance feedback evaluation, have a greater weight. Indeed, these latter are difficult to classify compared to patterns belonging to the majority class, negative. The results reach  a high level of correct classification. This indicates that the improvements over existing techniques can be associated with two aspects. The first involves the vector-based image representation, \textbf{KGEARSRG}, adopted. The second concerns AKS method for the classification stage. 


\unskip

\begin{table}[!ht]
\centering
\caption{Comparison results on the dataset
in~\cite{Haladova} and Table \ref{ds1}.}
\small
\begin{tabular}{|c|c|c|c|c|c|c|}
 \hline
\multicolumn{7}{|c|}{\emph{\textbf{AGF}}} \\
\hline
 \textbf{Problem} &
\textbf{AKS} & \textbf{C4.5} & \textbf{RIPPER}  & \textbf{L2-L
SVM} & \textbf{L2 RLR}& \textbf{RDR}\\ 
\hline
1 & 0.9414 & 0.5614 & 0.8234 & 0.6500 & 0.5456 & 0.8987\\
\hline
2 & 0.9356 & 0.8256 & 0.6600 & 0.8356 & 0.8078 & 0.7245\\
\hline
3 & 0.9678 & 0.8462 & 0.8651 & 0.4909 & 0.6123 & 0.7654\\
\hline

4 & 0.9746 & 0.8083 & 0.6600 & 0.4790 & 0.4104 & 0.6693\\
\hline
5 & 0.9654 & 0.7129 & 0.9861 & 0.8456 & 0.4432 & 0.6134\\

\hline
6 & 0.9342 & 0.5714 & 0.9525 & 0.8434 & 0.9525 & 0.5554\\
\hline
7 & 0.9567 & 0.6151 & 0.7423 & 0.5357 & 0.4799 & 0.6151\\
\hline
8 & 0.8345 & 0.4123 & 0.3563 & 0.7431 & 0.5124 & 0.7124\\
\hline
9 & 0.9435 & 0.9456 & 0.9456 & 0.8345 & 0.6600 & 0.6600\\
\hline
10 & 0.8456 & 0.4839 & 0.5345 & 0.4123 & 0.4009 & 0.5456\\
\hline
11 & 0.9457 & 0.9167 & 0.9088 & 0.9220 & 0.8666 & 0.9132\\
\hline
12 & 0.6028 & 0.5875 & 0.5239 & 0.4124 & 0.4934 & 0.5234\\
\hline
13 & 0.8847 & 0.7357 & 0.6836 & 0.7436 & 0.7013 & 0.5712\\
\hline
14 & 0.9376 & 0.9376 & 0.8562 & 0.8945 & 0.8722 & 0.8320\\
\hline

15 & 0.9765 & 0.8630 & 0.8897 & 0.8225 & 0.7440 & 0.8630\\
\hline
16 & 0.7142 & 0.5833 & 0.3893 & 0.4323 & 0.5455 & 0.5111\\
\hline
 \end{tabular}

\label{Res1}
\end{table}

\begin{table}[!ht]
\centering
\caption{Comparison results on the dataset
in~\cite{manzo2013attributed} and Table \ref{ds2}.}
\footnotesize
\begin{tabular}{|c|c|c|c|c|c|c|}
 \hline
\multicolumn{7}{|c|}{\emph{\textbf{AGF}}} \\
\hline
\textbf{Problem} &
\textbf{AKS} & \textbf{C4.5} & \textbf{RIPPER} & \textbf{L2-L
SVM} & \textbf{L2 RLR} & \textbf{RDR}\\
\hline
1 & 0.9822 & 0.6967 & 0.5122 & 0.4232 & 0.4322 & 0.6121\\
\hline
2 & 0.9143 & 0.5132 & 0.4323 & 0.4121 & 0.4212 & 0.5323\\
\hline
3 & 0.9641 & 0.4121 & 0.4211 & 0.4213 & 0.3221 & 0.4323\\
\hline
4 & 0.9454 & 0.4332 & 0.1888 & 0.4583 & 0.3810& 0.3810\\
\hline
5 & 0.9554 & 0.3810 & 0.2575 & 0.5595 & 0.3162 & 0.6967\\
\hline
6 & 0.9624 & 0.3001 & 0.1888 & 0.1312 & 0.3456 & 0.3121\\
\hline
7 & 0.9344 & 0.3810 & 0.5566 & 0.4122 & 0.4455 & 0.2234\\
\hline
8 & 0.9225 & 0.4333 & 0.1112 & 0.2575 & 0.1888& 0.1888\\
\hline
9 & 0.9443 & 0.6322 & 0.1888 & 0.1888 & 0.6122 & 0.6641\\
\hline
10 & 0.9653 & 0.1897 & 0.5234 & 0.6956 & 0.1888 & 0.1121\\
\hline
 \end{tabular}

\label{Res2}

\end{table}


\section{Conclusions}
\label{conc}

In this paper a structure for image representation called \textbf{Attributed Relational SIFT-based Regions Graph} (\textbf{ARSRG}) is presented through description and analysis of new aspects. Starting from previous works and performing a thorough study, theoretical notions have been introduced in order to clarify and deepen the structural design of \textbf{ARSRG}. It has been demonstrated how \textbf{ARSRG} can be adopted in disparate fields such as graph matching, graph embedding, bag of graph words and kernel graph embedding with application of object recognition and art painting retrieval/classification. The experimental results have amply shown how the performances on different datasets are better than state of art competitors. Future developments certainly include exploration of additional application fields, introduction of additional algorithms (mainly graph matching) to improve performance comparison and a greater enrichment of image features to include within the \textbf{ARSRG}.

\acks{This work is dedicated to Alfredo Petrosino. With him I took my first steps in the field of computer science. During these years spent together, I learned firmness in achieving goals, love and passion for the work. I will be forever grateful. Thank you my great master.}



\bibliography{sample}
\bibliographystyle{theapa}

\end{document}